\documentstyle[epsfig,12pt]{article}
\begin{document}

\begin{center}

{\bf
\title "The temperature dependence of equilibrium plasma density}

\bigskip

\bigskip
\author "B.V.Vasiliev
\bigskip

%{Институт физико-технических проблем,141980, Дубна, Россия}
Institute in Physical-Technical Problems, 141980, Dubna, Russia
\bigskip

{vasiliev@dubna.ru}
\end{center}

\bigskip
%\maketitle

\begin{abstract}
Temperature dependence of an electron-nuclear plasma equilibrium
density is considered basing on known approaches, which are given
in \cite{1}-\cite{2}. It is shown that at a very high temperature,
which is characteristic for a star interior, the equilibrium
plasma density is almost constant and  equals approximately to
$10^{25}$ particles per $cm^3$. At a relatively low temperature,
which is characteristic for star surface, the equilibrium plasma
density is in several orders lower and depends on temperature as
$T^{3/2}$.
\end{abstract}
\bigskip

\bigskip

\bigskip
\bigskip

The electron-nuclear plasma has a high density at high temperature
inside stars. It may be supposed  that a star exists in a steady
state, and the plasma is in a thermodynamic equilibrium. And so it
may be  expected that an equilibrium plasma should have a density
which is determined by its energy minimum.

It will be shown below that an energy minimum of equilibrium
plasma exists really. It enables to determinate an equilibrium
density of plasma as a function of its temperature. It is
important that the density of inner regions of stars is equal to
$a_0^{-3}\approx 10^{25}$ particles in $cm^3$ approximately ($a_0$
is Bohr radius) and it is in several orders lower for surface
regions, where it depends on temperature as $T^{3/2}$.

Let us consider an ensemble consisting of a permanent number of
particles N at constant temperature T. The equilibrium state at a
minimum of free energy F  will be settled by a ensemble volume V
change, i.e. by a change of a particle density n:

\begin{equation}
\biggl(\frac{\partial F}{\partial n}\biggr)_{N,T}=0.\label{eq}
\end{equation}

The chemical potential describes an assemble equilibrium. The
direct interaction between nuclei can be neglected in a dense
 electrically neutral equilibrium plasma. The plasma energy is
determined by electron-electron  interaction and electron-nucleus
interaction.

The chemical potential of an electron gas at high temperature (in
Boltzman approximation) is known \cite{1}:

\begin{equation}
\mu_B=kT~ln~\xi_B
\end{equation}

where

\begin{equation}
\xi_В=\frac{1}{2}\biggl[\frac{2\pi h^2}{m kT}
\biggr]^{3/2}n_e.\label{xi}
\end{equation}

and the electron gas density \cite{1}

\begin{equation}
n_e=\frac{2^{1/2}m^{3/2}}{\pi^2\hbar^3}\int_{0}^{\infty}\frac{\varepsilon^{1/2}
d\varepsilon}{e^{(\varepsilon
-\mu)/kT}+1}. \label{ne}
\end{equation}

We will search for the chemical potential of plasma at arbitrary
temperature as

\begin{equation}
\mu=kT~ln~\xi
\end{equation}

Using the substitution

\begin{equation}
\zeta=\frac{\varepsilon}{kT},
\end{equation}

from ({\ref{ne}}) we obtain

\begin{equation}
\xi=\xi_B\frac{\pi^{1/2}}{2I_1}
\end{equation}

where

\begin{equation}
I_1=\int_{0}^{\infty}\frac{\zeta^{1/2} d\zeta}{e^\zeta +\xi}
\end{equation}

Thus

\begin{equation}
\frac{d\xi_B}{d\xi}=\frac{2}{\pi^{1/2}}I_1\biggl(1-\xi\frac{I_3}{I_1}\biggr).
\end{equation}

where

\begin{equation}
I_3=\int_{0}^{\infty}\frac{\zeta^{1/2} d\zeta}{(e^\zeta +\xi)^2}.
\end{equation}

The free energy of an electron gas is  \cite{1}

\begin{equation}
F_e=\frac{N(2m)^{3/2}(kT)^{5/2}\xi}{3\pi^2\hbar^3
n_e}\int_{0}^{\infty}\frac{\zeta^{3/2} d\zeta}{e^\zeta
+\xi}=\frac{2I_2}{3I_1}NkT
\end{equation}

where

\begin{equation}
I_2=\int_{0}^{\infty}\frac{\zeta^{3/2} d\zeta}{e^\zeta +\xi}.
\end{equation}

At high temperature (in Boltzman approximation) when $\xi=\xi_B\ll
1$ and chemical potential $\mu=\mu_B<0$ and $|\mu_B|\gg 1$, the
free energy of an electron gas may be expanded in series. If we
conserve the two first terms of the series, we obtain

\begin{equation}
F_e=F_{ideal}+N\frac{\pi^{3/2}a_o^{3/2}
e^3}{4(kT)^{1/2}}n_e\label{Fe}
\end{equation}

Here the first term is the free energy of an ideal gas and the
second term is the correction for the identity of electrons. This
correction is positive as it takes into account that electrons can
not take places which are already occupied by other electrons and
it causes an increasing incompressibility of electron gas.

From Eq.({\ref{Fe}}) we obtain

\begin{eqnarray}
\biggl(\frac{dF_e}{dn_e}\biggr)_{N,T}=-\frac{2^{1/2}\pi^2
a_o^{3/2} e^3}{3(kT)^{1/2}}\frac{(I_4 I_1-I_3 I_2)}{I_1^3 (1-\xi
I_3/I_1)}
\end{eqnarray}

where $a_o$ is the Bohr radius  and

\begin{equation}
I_4=\int_{0}^{\infty}\frac{\zeta^{3/2} d\zeta}{(e^\zeta +\xi)^2}.
\end{equation}

In order to take into account a role of nuclei in the plasma
energy formation, we must calculate the correlation correction
\cite{2}:

\begin{eqnarray}
\delta F_{corr}=-N\frac{2\pi^{1/2}e^3
kT}{3}\biggl[Z^2\frac{n_i}{kT} +\biggl(\frac{\partial
n_e}{\partial \mu} \biggr)_{N,T}\biggr]^{3/2}\label{Fcor}
\end{eqnarray}

where Z is the nucleus charge, $n_i=n_e/Z$ is the nucleus density.

At high temperature (in Boltzman approximation) this correction is

\begin{equation}
\delta F_{corr}=-N\frac{2\pi^{1/2}e^3 (Z+1)^{3/2}}{3(kT)^{1/2}}
n_e^{1/2}
\end{equation}

This correction takes into account that nuclei "condense" the
electron gas in their closeness. On the contrary, the electron gas
density decreases away from the nuclei  and its compressibility
increases. This explains why this correction is negative.

Thus the free energy of plasma at high temperature with account
for both corrections

\begin{eqnarray}
F=F_{ideal}+N\frac{\pi^{3/2}a_o^{3/2} e^3}{4(kT)^{1/2}}n_e
-N\frac{2\pi^{1/2}e^3  (Z+1)^{3/2}}{3(kT)^{1/2}}n_e^{1/2}
\label{F}
\end{eqnarray}

i.e. the free energy of plasma has two correction added to ideal
value. They have different signs and differently depend on
particle density, but they equally depend on temperature. This
enables to calculate  the equilibrium density of a dense plasma at
high temperature from balance condition Eq.({\ref{eq}})

\begin{equation}
n_0=\frac{16 (Z+1)^3}{9\pi^2 a_o^3}\label{N0}
\end{equation}

 In this case the Fermi energy of electron gas  $\varepsilon_F
\approx \frac {Z^2e^2}{a_0}$ and the Boltzman approximation is
applicable when $kT\gg \varepsilon_F$. In this approximation the
equilibrium density does not depend on temperature.

For arbitrary temperature starting from  Eq.({\ref{Fcor}}), we
obtain

\begin{eqnarray}
\biggl(\frac{dF_{corr}}{dn_e}\biggr)_{N,T}=\frac{\pi^{1/2}
e^3(Z+1)^{3/2}}{3(kT n_e)^{1/2}} \biggl[1-\frac{\xi I_3}{(Z+1)
I_1} \biggr]^{3/2}\ast \nonumber \\ [2mm]
\ast\biggl[1-\frac{3\xi}{(Z+1)I_1}\frac{(I_3-2\xi I_5+\xi
I_3^2/I_1)}{(1-\xi I_3/I_1)[1-\frac{\xi I_3}{(Z+1) I_1}]}\biggr]
\end{eqnarray}

The equilibrium density of plasma is

\begin{eqnarray}
\frac{n_e}{n_0}=\biggl\{\frac{3}{2^{5/2}\pi^{1/2}}\frac{I_1^3(1-\xi
I_3/I_1)}{(I_4 I_1-I_2 I_3)}\biggl[1-\frac{\xi I_3}{(Z+1)
I_1}\biggr]^{3/2}\ast \nonumber
\\[2mm] \ast \biggl[1-\frac{3\xi}{(Z+1)I_1}\frac{(I_3-2\xi I_5+\xi
I_3^2/I_1)}{(1-\xi I_3/I_1)[1-\frac{\xi
I_3/I_1}{(Z+1)}]}\biggr]\biggr\}^2\label{nen0}
\end{eqnarray}

where

\begin{equation}
I_5=\int_{0}^{\infty}\frac{\zeta^{1/2} d\zeta}{(e^\zeta +\xi)^3}.
\end{equation}

At the same time

\begin{equation}
T=\frac{2\pi a_0 e^2}{k} \biggl(\frac{n_e}{\pi^{1/2}\xi I_1}
\biggr)^{2/3}\label{T}.
\end{equation}

The numerically calculated   dependence of the equilibrium plasma
density on temperature at Z=1 obtained from these equations is
shown in Fig.1.

Thus, at a high temperature $T\gg E_F/k$ the equilibrium density
of plasma approaches to $n_0$, and at a relatively low temperature
($T\approx E_F/k$)

\begin{equation}
{n_e}\sim T^{3/2}\label{nT1}.
\end{equation}

It is a consequence of that the full energy of electron gas in a
general case is \cite{1}

\begin{equation}
E_e=N\frac{2^{1/2} m^{3/2} (kT)^{5/2}\xi}{\pi^2\hbar^3
n_e}\int_{0}^{\infty}\frac{\zeta^{3/2} d\zeta}{e^\zeta
+\xi}\label{Ee}
\end{equation}

or with account for above formulas

\begin{equation}
E_e=NkT\frac{I_2}{I_1}\label{Ee1}.
\end{equation}

For a degenerate electron gas

\begin{equation}
E_e=\frac{3}{10}N E_F.
\end{equation}

As in this case the ratio $I_2/I_1\approx 1$, from
Eq.({\ref{Ee1}}) we obtain for the equilibrium degenerate plasma

\begin{equation}
E_F \approx kT.
\end{equation}

and with account for the definition of $E_F$, we obtain
Eq.({\ref{nT1}}).

The obtained equilibrium values of a plasma density enables to
explain a series of astrophysical effects: the proportionality of
star magnetic moments to their angular momenta (the Blackett
effect), the star mass spectrum and others \cite{3}.

\begin{figure}
\begin{center}
\includegraphics[8cm,1cm][14cm,11cm]{n(T).EPS}
%%\vspace{11cm}
\caption {The numerically calculated   dependence of the
equilibrium plasma density on temperature at Z=1.} \label{n(T)}
\end{center}
\end{figure}

\end{document}